\documentclass[
 aps, pra,
 amsmath,amssymb,
 11pt,
 final,
tightenlines,
 twoside,
 twocolumn,
 nofloats,
nofootinbib,
 superscriptaddress,
showkeys,
showkeywords,
 ]
{revtex4-2}

\usepackage[T2A]{fontenc}
\usepackage[utf8]{inputenc}
\usepackage[english]{babel}
\usepackage{graphicx}
\usepackage{dcolumn}
\usepackage{bm}

\usepackage{xcolor}
\usepackage{soul}
\usepackage{changes}

\input{maik.rty}

\setcitestyle{authoryear,round}
\def\squareforqed{\hbox{\rlap{$\sqcap$}$\sqcup$}}

\def\sq{\ifmmode\squareforqed\else{\unskip\nobreak\hfil
\penalty50\hskip1em\null\nobreak\hfil\squareforqed
\parfillskip=0pt\finalhyphendemerits=0\endgraf}\fi}

\def\ga{\mathrel{\mathchoice {\vcenter{\offinterlineskip\halign{\hfil
$\displaystyle##$\hfil\cr>\cr\sim\cr}}}
{\vcenter{\offinterlineskip\halign{\hfil$\textstyle##$\hfil\cr
>\cr\sim\cr}}}
{\vcenter{\offinterlineskip\halign{\hfil$\scriptstyle##$\hfil\cr
>\cr\sim\cr}}}
{\vcenter{\offinterlineskip\halign{\hfil$\scriptscriptstyle##$\hfil\cr
>\cr\sim\cr}}}}}

\def\degr{\hbox{$^\circ$}}

\def\arcsec{\hbox{$^{\prime\prime}$}}

\def\utw{\smash{\rlap{\lower5pt\hbox{$\sim$}}}}

\def\udtw{\smash{\rlap{\lower6pt\hbox{$\approx$}}}}

\def\fm{\hbox{$\,.\!\!^{\rm m}$}}

\def\diameter{{\ifmmode\mathchoice
{\ooalign{\hfil\hbox{$\displaystyle/$}\hfil\crcr
{\hbox{$\displaystyle\mathchar"20D$}}}}
{\ooalign{\hfil\hbox{$\textstyle/$}\hfil\crcr
{\hbox{$\textstyle\mathchar"20D$}}}}
{\ooalign{\hfil\hbox{$\scriptstyle/$}\hfil\crcr
{\hbox{$\scriptstyle\mathchar"20D$}}}}
{\ooalign{\hfil\hbox{$\scriptscriptstyle/$}\hfil\crcr
{\hbox{$\scriptscriptstyle\mathchar"20D$}}}}
\else{\ooalign{\hfil/\hfil\crcr\mathhexbox20D}}%
\fi}}

\begin{document}




\title{ZZ~Tau~IRS: a low mass UX~Ori type star with strong wind
}

\author{\firstname{M.~A.}~\surname{Burlak}}
\author{\firstname{A.~V.}~\surname{Dodin}}
\author{\firstname{A.~V.}~\surname{Zharova}}
\author{\firstname{S.~G.}~\surname{Zheltoukhov}}
\author{\firstname{N.~P.}~\surname{Ikonnikova}}
\author{\firstname{S.~A.}~\surname{Lamzin}}
\email{lamzin@sai.msu.ru}
\author{\firstname{S.~A.}~\surname{Potanin}}
\author{\firstname{B.~S.}~\surname{Safonov}}
\author{\firstname{I.~A.}~\surname{Strakhov}}
\author{\firstname{A.~M.}~\surname{Tatarnikov}}

\affiliation{Sternberg Astronomical Institute, M.V.~Lomonosov Moscow State University, Moscow, 119234 Russia\\}


\begin{abstract}
The results of photometric, polarimetric and spectroscopic observations of the young star ZZ~Tau~IRS in the visible and near-infrared bands are presented. Against the continuum of an M spectral type star about 50 emission lines of allowed (H\,I, He\,I, Na\,I, S\,II) and forbidden (O\,I, O\,II, O\,III, N\,I, N\,II, S\,II, Ca\,II, Fe\,II, Ni\,II) transitions were identified. It was found that from the autumn of 2020 to the beginning of 2023, the brightness of the star in the visible region decreased $(\Delta I \approx 1.5^m),$ and then began to return to initial level. As the visible brightness of the star declined, its colour indices decreased in the visible region, but increased in the near-IR bands. At light minimum, the degree of polarization in the $I$ band reached $\approx 13$\,\%{}, and the equivalent widths of e.g. the H$\alpha$ and [S\,II]~6731~\AA{} lines increased to 376 and 79~\AA{}, respectively. Arguments are given in favour of ZZ~Tau~IRS being a UX~Ori type star, and its variability being due to eclipses by dust clouds, which are inhomogeneities in the dusty disc wind. Forbidden lines are formed both in the disc wind and in the jet, the axis of which is oriented along $\text{PA}=61\pm 3\degr.$ The jet mass-loss rate is $>5 \times 10^{-10}$~M$_\odot$/yr, what is abnormally large for a star with a mass of $<0.3$~M$_\odot.$ Apparently, the disc wind of ZZ~Tau~IRS is not axially symmetric, probably due to the azimuthal asymmetry of the protoplanetary disc found earlier from ALMA observations.
\end{abstract}

Accepted by {\it Astrophysical Bulletin}
 
\maketitle

\section{Introduction}
\label{sect:introduct}

\citet{Strom-1989} concluded that the optical counterpart of the far-infrared (FIR) source IRAS~04278$+$2435 \citep{Rucinski-1985} is a young star which lies $\approx 35\arcsec$ south of ZZ~Tau. \citet{KH-1995} found that FIR luminosity of the star is larger than 50\,\% of its bolometric luminosity and on this basis named the star ZZ~Tau~IRS.

The {\it Gaia} parallax for ZZ~Tau~IRS ({\it Gaia} EDR3 id147869573608324992) is $9.46 \pm 0.61$~mas, which corresponds to a distance of $103.7 \pm 6.7$~pc \citep{Gaia-16b, Gaia-2020}. However, as was mentioned by \citet{Hashimoto-2021}, the error of the astrometric solution is large ($\text{RUWE}=2.49$), so we will further use the distance $d\approx 130$~pc adopted from \citet{Akeson-2019}.

The first (and thus far the only one) high resolution optical (0.635-0.874~$\mu m$) spectrum of this faint 
\footnote{
According to {\lq}The Second-Generation Guide Star Catalog{\rq}
\citep{Lasker-2008} $V=18.17 \pm 0.39,$ $B=18.70 \pm 0.39,$ but only average epoch (t=1994.776) of these (at least two) observations is presented in the Catalog.}
star was observed and analysed by \citet{White-Hillenbrand-2004}. They concluded that ZZ~Tau~IRS is an M$4.5 \pm 2$ very reddened $(A_{\text V}=7.6)$ classical T~Tauri star (CTTS) with unusually strong emission lines: e.g. the equivalent widths (EW) of the H$\alpha$ and [S\,II] 6731 lines are 238 and 77~\AA{}, respectively. To explain these features the authors suggested that ZZ~Tau~IRS has a protoplanetary disc which is seen close to edge-on.

The low-resolution ($R\sim 3\,000$) spectroscopic observations by \citet{Guieu-2006} confirmed this conclusion. The authors found that the star has a spectral type of M5.25 ($T_{\text eff}=3100$~K), but is not so much reddened: $A_{\text V}=2.4.$ Later, \citet{Herzeg-Hillebrandt-2014} found an even smaller value of extinction $A_{\text V}=1.7,$ but practically the same effective temperature as \citet{Kounkel-2019}: $T_{\text eff}=3077 \pm 32$~K. Despite such a low $T_{\text eff}$ and a fairly large $A_{\text V}$, the star was observed in the near $(\lambda_{\text eff}=0.23~\mu m$) and far $(\lambda_{\text eff}=0.153~\mu m$) ultraviolet (UV) bands by the {GALEX} Space Observatory: $m_{\text {FUV}}=21.02\pm 0.23$ and $m_{\text {NUV}}=21.53 \pm 0.29$ \citep{GALLEX-2017}.

ALMA observations by \citet{Hashimoto-2021} revealed a protoplanetary disc around ZZ~Tau~IRS inclined at $\approx 60\degr$ to the line of sight with a position angle of the disc major axis of ${\text PA}\approx 135\degr$. \citet{Hashimoto-2021} also concluded that the mass of the star is $0.1-0.3$~M$_\odot$ which is in agreement with the value $M=0.1-0.2$~M$_\odot$ found from spectroscopic optical observations \citep{Andrews-2013, Herzeg-Hillebrandt-2014}.
 
\citet{Gomez-1997} detected diffuse [S\,II] emission, presumably the Herbig-Haro object HH~393, located $28\arcsec$ south-west of ZZ~Tau~IRS. Later, two filaments were also found in the vicinity of ZZ~Tau~IRS. The first one, located north-west of the star, was observed by \citet{Hodapp-1994} in the $K^\prime$ spectral band, which included the H$_2$ ($\lambda=2.12$~$\mu m$) and Br$_\gamma$ emission lines, and the second one was observed by \citet {Bally-2012} in H$\alpha$ south-west of ZZ~Tau~IRS. Physical parameters and kinematics of the filaments as well as those of HH~393 were considered by \citet{Dodin-2023}.

The rest of the paper is organized in a standard way. In Section~\ref{sect:observation} we describe our observations, in Section~\ref{sect:results} we talk about our results, and discuss them in Section~\ref{sect:discuss}. Finally, we summarize our conclusions in Section~\ref{sect:concludion}.
                 
\section{Observations}
 \label{sect:observation}

Optical photometry of ZZ~Tau~IRS was carried out with the 0.6-m telescope of the Caucasian Mountain Observatory (CMO) of the Sternberg Astronomical Institute of the Lomonosov Moscow State University (SAI MSU) equipped with a CCD camera and a set of standard Bessel-Cousins $BVRI$ filters \citep{Berdnikov2020}. Respective magnitudes of comparison stars were adopted from AAVSO\footnote{https://www.aavso.org}.
We should mention that our $B$ magnitudes are somewhat brighter due to the so-called {\lq}red leakage{\rq} \citep{Nikishev_2023}, which is difficult to take into account for such a faint red star.

We found no information in literature about photometric observations of ZZ~Tau~IRS in the visible band (except those mentioned in Introduction) and tried to find the star in photographic plates from the SAI Astronomical Plate Stack which were obtained from 1965 March 26 till 1988 February 15. About 300 plates centered at RY~Tau with the photometric system close to the Johnson $B$ band were examined. The limiting magnitude varies between $B\approx 17-18^m$, and ZZ~Tau~IRS was found in 9 of these plates -- see Table~\ref{tab:B-lc}. Note that hereinafter we use reduced Julian Date ${\text rJD}={\text JD}-2\,400\,000.$

\begin{table}
\renewcommand{\tabcolsep}{0.2cm}
 \caption{Photographic magnitudes}
  \label{tab:B-lc}
 \begin{tabular}{lll||lll}
\hline
rJD & $m_{\text pg}$ & $\sigma_{\text m}$ & rJD & $m_{\text pg}$ & $\sigma_{\text m}$ \\
  \hline
43\,865.335 & 18.2 & 0.3 & 46\,763.322 & 18.8 & 0.5  \\
43\,379.420 & 18.4 & 0.5 & 46\,771.384 & 17.9 & 0.5  \\
43\,406.422 & 18.4 & 0.5 & 47\,200.200 & 17.1 & 0.2  \\
43\,466.282 & 17.9 & 0.4 & 47\,207.237 & 17.7 & 0.3  \\
46\,497.248 & 17.7 & 0.4 &             &      &      \\
  \hline
 \end{tabular} \\
\end{table}

Near-infrared (NIR) observations of ZZ~Tau~IRS were carried out in the $YJHK$ bands of the MKO-NIR photometric system at the 2.5-m telescope of CMO SAI MSU equipped with the IR camera-spectrograph ASTRONIRCAM \citep{Nadjip-17}. The details of observations and data reduction are described in \citet{Tatarnikov-2023}. The results of our observations are presented in Table~\ref{tab:tab2} along with observational data found in literature.

\begin{table*}
\renewcommand{\tabcolsep}{0.08cm}
\caption{NIR photometry of ZZ~Tau~IRS} 
 \label{tab:tab2}
\begin{tabular}{ccccccc||ccccccccc}
\hline
rJD   & $J$ & $\sigma_J$ & $H$ & $\sigma_H$ & $K$ & $\sigma_K$ & 
rJD   & $Y$   & $\sigma_Y$ & $J$ & $\sigma_J$ & $H$ & $\sigma_H$ & $K$ & $\sigma_K$ \\
\hline
48\,562.75$^a$ & 10.33 & 0.02 &  9.37 & 0.02 &  8.72 & 0.03 &
 59\,944.32$^c$ &       &      & 14.28 & 0.01 & 12.61 & 0.01 & 10.84 & 0.01 \\
50\,782.77$^b$ & 12.84 & 0.02 & 11.44 & 0.03 & 10.31 & 0.02 &  
 59\,953.45$^c$ & 15.32 & 0.02 & 14.41 & 0.05 & 12.85 & 0.08 & 10.96 & 0.04 \\
59\,079.53$^c$ & 13.25 & 0.01 & 12.32 & 0.01 & 11.46 & 0.01 &
 59\,954.35$^c$ & 15.35 & 0.02 & 14.43 & 0.02 & 12.77 & 0.06 & 10.95 & 0.03 \\
59\,084.49$^c$ & 13.18 & 0.01 & 12.23 & 0.01 & 11.33 & 0.02 &
 59\,966.15$^c$ & 15.09 & 0.03 & 14.10 & 0.02 & 12.37 & 0.05 & 10.42 & 0.05 \\
59\,091.57$^c$ & 13.19 & 0.01 & 12.26 & 0.01 & 11.40 & 0.01 &
 59\,976.24$^c$ & 15.06 & 0.06 & 99.99 & 9.99 & 12.70 & 0.04 & 10.79 & 0.03 \\ 
59\,094.49$^c$ & 13.21 & 0.01 & 12.29 & 0.02 & 11.43 & 0.02 &
 59\,985.20$^c$ & 15.05 & 0.01 & 14.20 & 0.10 & 12.82 & 0.05 & 11.05 & 0.04 \\ 
59\,551.26$^c$ & 13.08 & 0.07 & 12.09 & 0.02 & 11.24 & 0.02 &
 60\,000.17$^c$ & 15.02 & 0.03 & 14.11 & 0.02 & 12.68 & 0.05 & 10.94 & 0.03 \\ 
59\,899.24$^c$ & 14.15 & 0.03 & 12.53 & 0.03 & 10.91 & 0.02 &
 60\,012.32$^c$ & 15.43 & 0.02 & 14.57 & 0.02 & 13.01 & 0.01 & 10.83 & 0.01 \\ 
59\,907.34$^c$ & 13.69 & 0.04 & 11.97 & 0.01 & 10.40 & 0.01 &
 60\,188.52$^c$ & 14.49 & 0.03 & 13.82 & 0.01 & 12.69 & 0.02 & 11.35 & 0.03 \\ 
59\,915.40$^c$ & 13.57 & 0.08 & 11.88 & 0.07 & 10.34 & 0.07 &
 60\,202.50$^c$ & 14.42 & 0.01 & 13.57 & 0.01 & 12.22 & 0.03 & 10.83 & 0.02 \\ 
59\,920.32$^c$ & 13.53 & 0.02 & 11.84 & 0.02 & 10.29 & 0.01 &
             &       &      &       &      &       &      &       &       \\
\hline
\end{tabular} \\
a) \citet{KH-1995}; b) -- \citet{Cohen-2003}; c) -- this work.
\end{table*}

In February 2023, a new infrared LMP camera \footnote{$L$ and $M$ Photometer, based on the Gavin615A detector \citep{Zheltoukhov-2022}} 
was mounted on the 2.5~m telescope of CMO, and we used it to estimate the brightness of the star in the $L$ $(\lambda_c=3.7~\mu m, \Delta \lambda_{0.5}=0.49~\mu m)$ and $M$ $(\lambda_c=4.8~\mu m, \Delta \lambda_{0.5}=0.54~\mu m)$ bands.

Polarimetric observations of ZZ~Tau~IRS were carried out in the $I_{\text c}$ band with the SPeckle Polarimeter (SPP) on the 2.5-m telescope of CMO SAI MSU \citep{Safonov-17}. The details of observations and data reduction were described by \citet{Dodin-19}, the results are presented in Table~\ref{tab:polarim}.

\begin{table}
\renewcommand{\tabcolsep}{0.15cm}
\caption{Polarimetry of ZZ~Tau~IRS}
 \label{tab:polarim} 
\begin{tabular}{lllll} 
\hline 
rJD  & $p$ & $\sigma_{\text{p}}$ & $\theta$ & $\sigma_{\theta}$ \\
     & \%  & \%               & $\degr$  & $\degr$           \\ 
\hline 
59\,248.31 & 11.33 & 0.22  & 163.5 & 0.6 \\
59\,305.21 & 11.12 & 0.34  & 163.8 & 0.9 \\
59\,517.42 & 11.62 & 0.27  & 155.1 & 0.7 \\  
59\,628.27 & 11.14 & 0.38  & 165.0 & 2.0 \\
59\,914.14 & 13.21 & 0.67  & 180.6 & 2.9 \\
60\,202.55 & 10.3  & 0.4   & 146.0 & 1.0 \\
\hline 
\end{tabular}\\ 
Cols\,2--3: the polarization degree and its error;
Cols\,4--5: the polarization angle and its error.
\end{table}

Spectroscopic data were obtained with the Transient Double-beam Spectrograph (TDS) -- see \citet{Potanin-2020} for the description of the instrument and the data reduction procedures. The spectral resolution of the TDS is $R=\lambda/\Delta \lambda \approx 2400$ in the red channel and $\approx 1300$ in the blue one (FWHM$\approx 120$ and 240~km~s$^{-1},$ respectively see fig. \,3 in \citealt{Belinski-2023}) when a $1\arcsec$ slit is used. The spectra were wavelength-calibrated using the spectrum of an emission-line lamp and then a correction based on 35-38 emission sky lines was applied. For the red channel, the residual scatter of the positions of the sky lines is 3–4 km~s$^{-1}$ and this constrains the wavelength-calibration accuracy of the calibration in the red channel. The accuracy was worse in the blue channel due to a small number of telluric lines. For the $\lambda \gtrsim 5000$~\AA{} region where the [O\,I]~5577~\AA{} line was the only reference point the accuracy is $\approx 10$~km/s. 

We also made use of high-resolution spectra of ZZ~Tau~IRS retrieved from the KOA archive of the Keck/HIRES spectra (spectral resolution $R\approx 36\,000,$ PI: R.~White)\footnote{https://koa.ipac.caltech.edu/cgi-bin/KOA/nph-KOAlogin}.
All the used Keck spectra are of {\lq}scientific grade{\rq}, so we did not process them additionally. Information about spectroscopic observations is given in Table~\ref{tab:spectra}.

\begin{table}
\renewcommand{\tabcolsep}{0.12cm}
 \caption{Spectral data} 
  \label{tab:spectra}
\begin{tabular}{lclll}
\hline
rJD & Spectrograph   & $\Delta \lambda$, nm  & $\Delta t,$ min & d, \arcsec \\
\hline
52\,688.75 & HIRES  & 624-868 & $3\times 15$ & 1.2  \\
59\,503.55 & TDS  & 370-745 & $8\times 10$ & 1.0 \\
59\,504.54 & TDS  & 370-745 & $8\times 10$ & 1.0 \\
59\,516.54 & TDS  & 370-745 & $8\times 10$ & 1.0 \\
59\,955.27 & TDS  & 370-745 & $8\times 20$ & 1.5 \\
60\,203.56 & TDS  & 370-745 & $3\times 10$ & 1.5 \\
\hline
  \end{tabular} \\
$\Delta t$ -- exposure time, $d$ -- slit width.
\end{table}

In our study, we also used direct images of the region around ZZ~Tau~IRS obtained with the NBI CCD camera of the 2.5-m telescope\footnote{https://obs.sai.msu.ru/cmo/sai25/wfi/} in two filters: [S\,II]~$6716 + 6731$~\AA{} and nearby continuum [S\,II]rc -- see \citet{Dodin-2023} for details.

One more image of ZZ~Tau~IRS we used was taken from the Hubble Space Telescope (HST) archive. It was obtained on 2017 August 8 in  the F160W filter $(\lambda_0 = 1.55~\mu m, FWHM=0.3~\mu m$) with the WFC3/IR camera\footnote{ PI: T. Megeath, Proposal ID:14181, https://mast.stsci.edu}. 
To increase the contrast of the image and eliminate diffraction {\lq}rays{\rq}, we subtracted from the original image a point source model based on 3 nearest bright stars, and applied smoothing with a two-dimensional Gaussian function.

\section{Results}
\label{sect:results}

\subsection{Photometric and polarimetric data}
 \label{sect:photopolarimetry}

%
\begin{figure}
 \begin{center}
\includegraphics[width=\columnwidth]{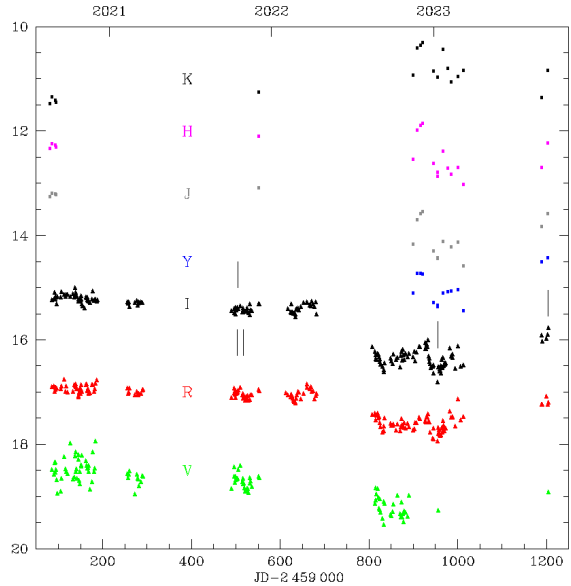}
 \end{center}
  \caption{$VRIYJHK$ light curves of ZZ~Tau~IRS. Squares are individual observations, and triangles represent nightly averaged magnitudes. Vertical lines mark moments of our spectroscopic observations.}  
 \label{fig:lcurve}
\end{figure}
The light curves of ZZ~Tau~IRS in the visible and NIR photometric bands based on our observations are presented in Fig.~\ref{fig:lcurve}, so that the $VRI$ observations are averaged over each night to increase signal to noise (S/N) ratio. As can be seen, brightness variations display the maximum amplitude in the $I$ band. From August 2020 to January 2022 ({\lq}bright{\rq} state) the mean $I$-band brightness of the star gradually decreased from $I\approx 15\fm 1$ by nearly $0\fm 2$ and then grew by $\approx 0\fm 1$ in January--April 2022. But in the beginning of the next season of visibility (August 2022) the star was found $\sim 0.6^m$ fainter than in April. Later (up to March 2023) the brightness oscillated around an average value of $I\approx 16\fm 3$ ({\lq}faint{\rq} state), but in late August 2023 the star apparently began to return to the {\lq}bright{\rq} state.

%
\begin{figure}
 \begin{center}
\includegraphics[width=\columnwidth]{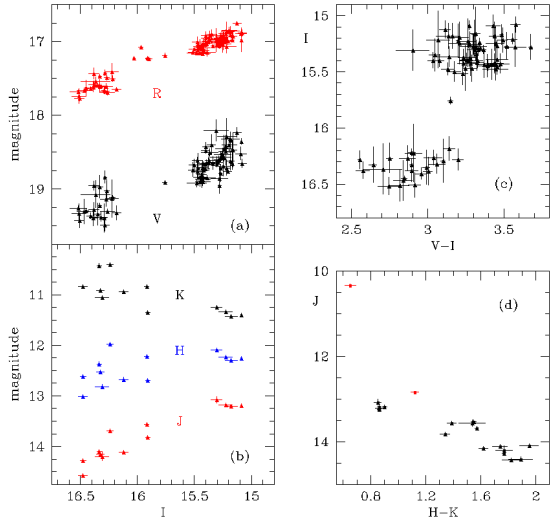}
 \end{center}
  \caption{Variations of ZZ~Tau~IRS brightness in the $R,$ $V$ (panel a) and $J,$ $H,$ $K$ (panel b) bands versus those in the $I$ band. Only data observed on the same night are presented. Colour--magnitude diagrams $I$ vs. $V-I$ (panel c) and $J$ vs. $H-K$ (panel d) are also shown. Black points in panel d are our observations, and red triangles represent the data of \citet{KH-1995}, \citet{Cohen-2003}, presented in Table~\ref{tab:tab2}.
  }
 \label{fig:colors}
\end{figure}
%

As follows from Fig.~\ref{fig:colors}a, the $R$ and $V$ brightness varied in a similar way, with the star bluer when fainter (Fig.~\ref{fig:colors}c). The $R-I$ colour changed similarly despite the fact that the contribution of emission lines, especially H$\alpha$, to the $R$ band increased as the star faded (see Sect.~\ref{sect:activity}). 
 
\begin{figure}
 \begin{center}
\includegraphics[width=\columnwidth]{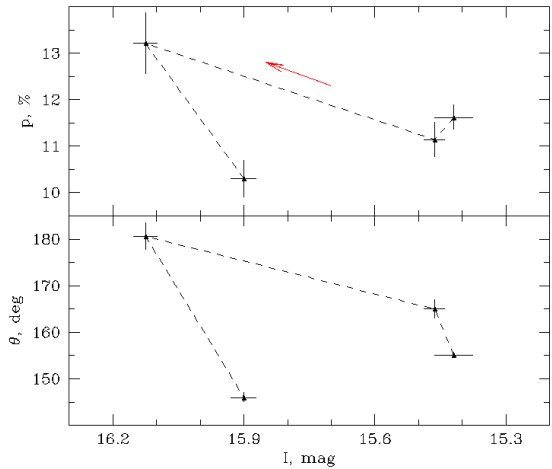}
 \end{center}
  \caption{Plot showing the change of the polarization degree (upper panel) and polarimertic angle (bottom panel) with the change of the $I$-band brightness of the star.}
 \label{fig:polariz-vs-mag}
\end{figure}

We suppose that the photometric behaviour of ZZ~Tau~IRS in the visible region is likely due to the same reason as in the case of UXORs (UX~Ori type stars): as the star fades, the contribution of bluer light, scattered by the circumstellar dust, increases \citep{Grinin-88}. A further argument in favour of this interpretation is an increase in polarization during the transition of the star from the {\lq}bright{\rq} to {\lq}faint{\rq} state -- see Fig.~\ref{fig:polariz-vs-mag}.

%

\begin{figure}
 \begin{center}
\includegraphics[width=\columnwidth]{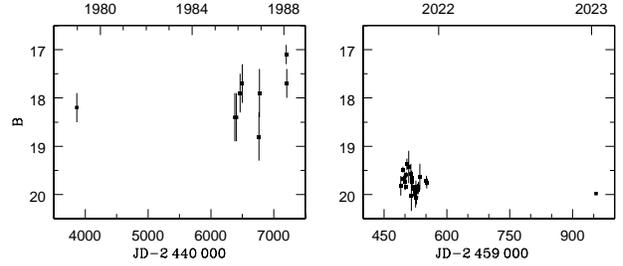}
 \end{center}
  \caption{Historical light curve of ZZ~Tau~IRS, based on our photographic (left panel) and CCD (right panel) observations.}
 \label{fig:B-band-lc}
\end{figure}

At the initial stage of fading, an UXOR-type star usually becomes redder, and only when its brightness has decreased significantly, it starts to get bluer \citep{Wenzel-1969}. In the case of ZZ~Tau~IRS, the absence of the {\lq}reddening stage{\rq} probably means, that even in 2020 the scattered light dominated in the visible band. This conclusion is supported by the fact that the star was noticeably brighter in the $B$ band in the 80s of the last century than in 2020-2021, as follows from Fig.~\ref{fig:B-band-lc} showing our photometric observations and from Table~\ref{tab:B-lc}. Besides, the polarization degree $p_I$ of the star in the {\lq}bright state{\rq} was larger then 11~\%{} (see Table~\ref{tab:polarim}) -- as far as we know, a larger value of $p_I$ in UXOR-type stars was observed only for RW~Aur~A \citep{Dodin-19}.

Note also that the polarization angle of ZZ~Tau~IRS in the {\lq}bright{\rq} state $\theta \approx 160 \degr$ is close to the position angle of the major axis of its circumstellar disc $PA_{\text{disk}}=135 \degr$ \citep{Hashimoto-2021}. Such a polarimetric angle is expected if the light is scattered by the circumstellar dust extended along the rotation axis of the disc rather than by the disc itself \citep{Whitney-93}.

%

As can be seen from Fig.~\ref{fig:colors}b, a gradual visual brightness decreasing in 2020-2022 was accompanied by a small brightening in the $JHK$ bands. After the star had transitioned to the {\lq}faint{\rq} state in the visible region, the $J$ and $H$-band brightness on the contrary reached maximum value and then decreased, but the $K$-band brightness continued to increase. Throughout the entire period of our observations, the brighter the star in the NIR region, the {\lq}redder{\rq} its $H-K$ (see Fig.~\ref{fig:colors}d) and $J-H$ colour indices.

According to Table~\ref{tab:tab2}, the star was much brighter in the NIR region thirty years ago: for example, it was brighter in the $K$ band by $\approx 2.5$~mag on 1991 November 2 \citep{KH-1995} and by $\approx 1$~mag on 1997 November 3 \citep{Cohen-2003} than in 2020. Nevertheless, the pattern of NIR brightness variations was the same as during our observations: the $H-K$ (see Fig.~\ref{fig:colors}d) and $J-H$ colours being redder when the $J$ brightness fainter. 

%
\begin{figure}
\includegraphics[width=\columnwidth]{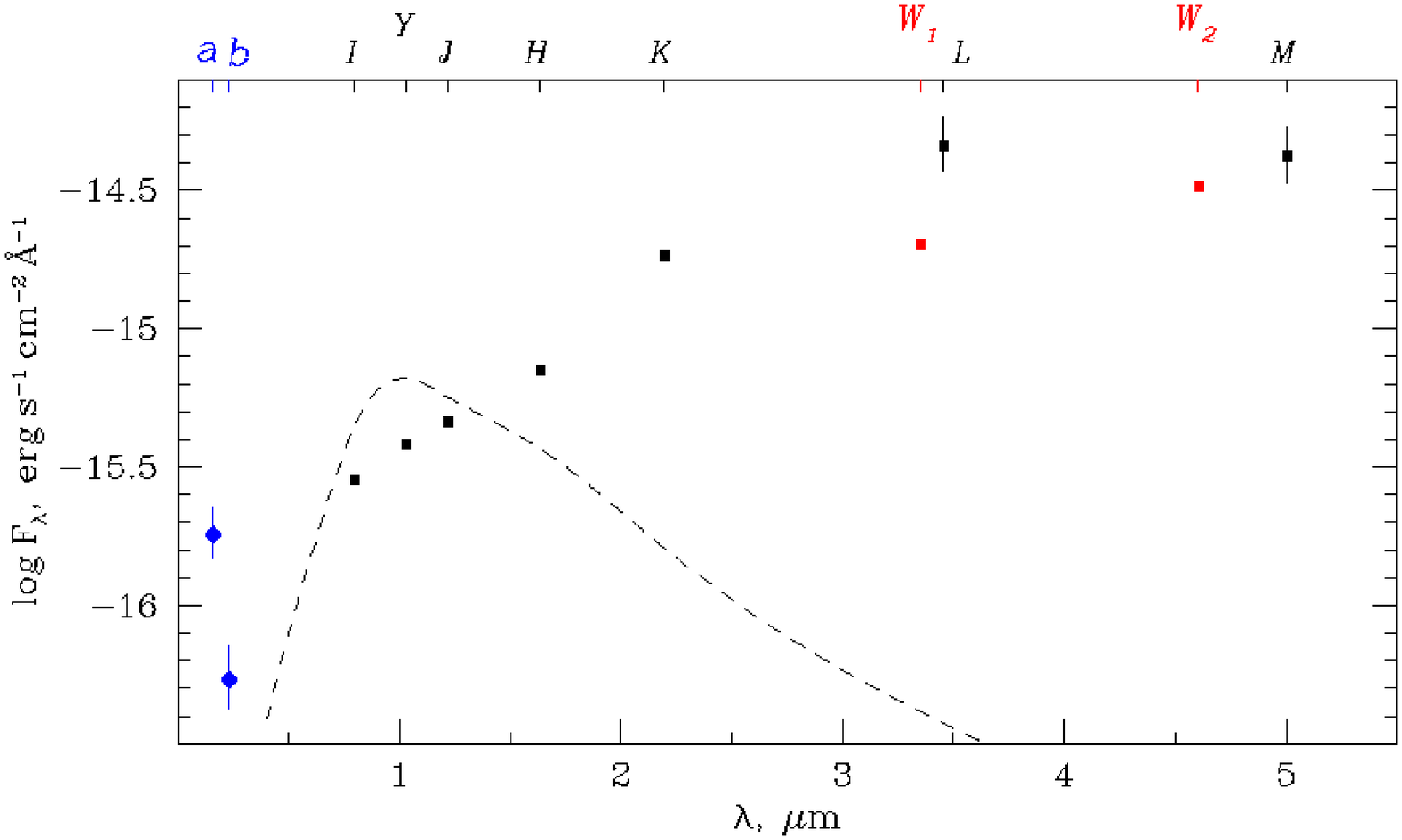}
\caption{Comparison of the spectral energy distribution (SED) of ZZ~Tau~IRS obtained on 2023 March 8 (black squares) with the data from {WISE} (red squares) and {GALEX} (blue circles) space observatories. The symbols $a$ and $b$ correspond to the GALEX near and far UV magnitudes, respectively. Dashed line is the theoretical SED of a star with $T_{\text eff} = 3100$~K and $\lg g=4.0,$ based on synthetic colour indices from the BT-Settl model \citep{Pecaut-Mamajek-2013} normalised to $I=16\fm0$.}
 \label{fig:sed}
\end{figure}
%

Thus, ZZ~Tau~IRS was significantly brighter both in the visible and NIR bands before 2020 (at least sometimes) which is in agreement with the assumption that by the time of our observations the star was in an deep enough eclipse by circumstellar dust. 

On 2023 March 8, we managed to observe ZZ~Tau~IRS not only in the $JHK$ but also $LM$ bands. Comparing the obtained values $L=7.96 \pm 0.24,$ $M=6.78 \pm 0.25$ with the observations carried out in February 2010 by the {WISE} space observatory \citep{Wright-10} in the $W_1$ $(\lambda_{\text eff}=3.35~\mu m$) and $W_2$ $(\lambda_{\text eff}=4.60~\mu m$) filters\footnote{W$_1=9.00\pm 0.02,$ W$_2=7.15\pm 0.02$} 
indicates that the flux at $3-5~\mu m$ was larger in 2023 near the visible light minimum than in 2010 (see Fig.~\ref{fig:sed}).


\subsection{Spectroscopic observations}
 \label{sect:star-spectra}

According to \citet{White-Hillenbrand-2004} and \citet{Kounkel-2019} radial velocity of ZZ~Tau~IRS is $V_{\text r}=18.4 \pm 3.7$ and $18.6 \pm 1.9$~km/s, respectively, so we further use $V_{\text r}=18.5$~km/s.

Since we analyzed the same high-resolution {Keck} spectra of the star as \citet{White-Hillenbrand-2004}, we found the same EWs of the most prominent emission lines, e.g. EW$_{H\alpha} \approx 240$~\AA{}. In addition to the [O\,I]~6363~\AA{}, H$\alpha,$ [N\,II] $6548+6583$~\AA{}, [S\,II] $6717+6731$~\AA{}, Ca\,II~$8498+8662$~\AA{} lines mentioned by \citet{White-Hillenbrand-2004}, we identified in these spectra the He\,I~6678~\AA{}, [Fe\,II]~7155~\AA{} and [Ni\,II]~7378~\AA{} emission lines. 

\begin{figure}
 \begin{center}
\includegraphics[width=\columnwidth]{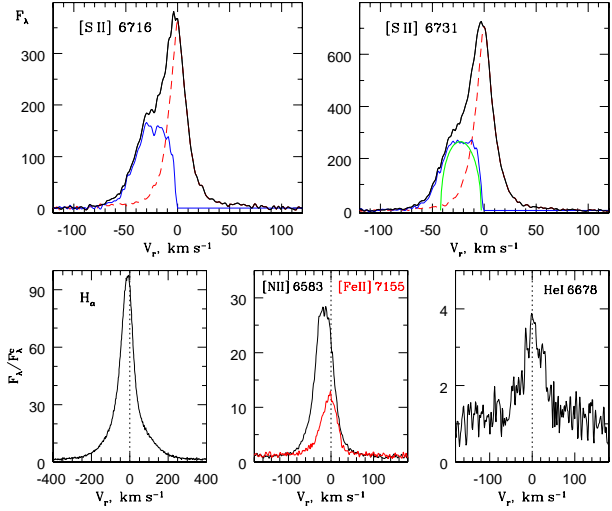}
 \end{center}
  \caption{Profiles of some emission lines from the high-resolution spectrum of ZZ~Tau~IRS in the star rest frame. Upper panels -- profile decomposition of [S\,II]~6717 and 6731~\AA{} lines into LV and HV components. The theoretical profile described in Sect.~\ref{sect:jet-accretion-params}, is shown with bold green line. The $F_\lambda$ flux along vertical axis is in units of $10^{-17}$~erg~s$^{-1}$~cm$^{-2}$~\AA$^{-1}.$ The line profiles in the bottom panels are normalised to continuum level.}
 \label{fig:keck-profs}
\end{figure}

As is seen in Fig.~\ref{fig:keck-profs} all emission line profiles, except for He\,I~6678~\AA{}, appear asymmetric and can be represented by a sum of two components: a {\lq}low velocity{\rq} (LV) component, nearly symmetric relative to the star rest frame\footnote{More exactly, LV components are also blue-shifted relative to the star rest frame by 2-3 km/s.}
and a {\lq}high velocity{\rq} (HV) component, blue-shifted by $\approx 30$~km~s$^{-1}$. The LV components are located in the velocity range from approximately $-60$ to $60$~km~s$^{-1},$ and the blue wing of HV component extends up to $\approx -60$~km~s$^{-1}.$ The relative contribution $f_{\text b}$ of the HV component flux to the total line flux is different for different lines: e.g., $f_{\text b}\approx 0.33$ and 0.22 in the case of [S\,II]~6717 and [Fe\,II]~7155 lines, respectively.

It seems natural to interpret the observed structure of emission lines in terms of the two-component outflow model typical for CTTSs \citep{Kwan-Tademaru-1988, Kwan-Tademaru-1995}, which assumes that the HV component originates in the collimated flow (jet) and the LV component is formed in the poorly collimated disc wind, which in our case contains dust and is responsible for the UXOR-like variability of the star.

\begin{figure}
\includegraphics[width=\columnwidth]{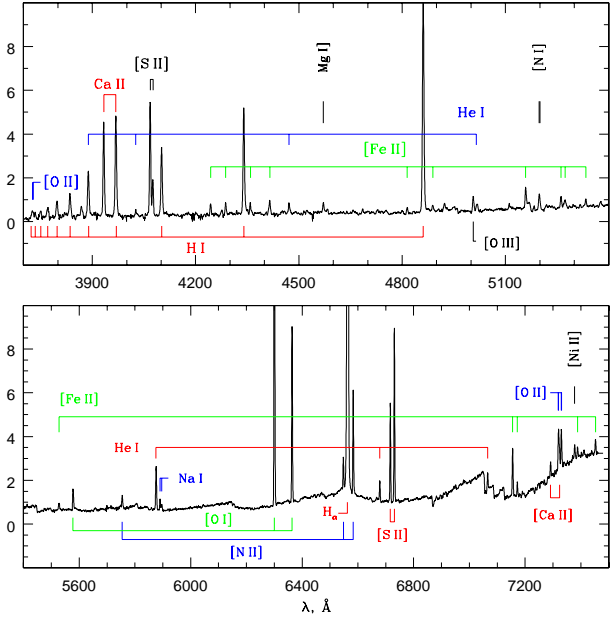}
\caption{An average of three spectra of ZZ~Tau~IRS obtained during the {\lq}bright state{\rq} of the star. The monochromatic flux $F_\lambda$ is plotted along vertical axis in arbitrary units.}
 \label{fig:tds-average-spectrum}
\end{figure}

Our low-resolution spectra reveal much more emission lines. As can be seen from Fig.~\ref{fig:tds-average-spectrum}, we identified about two tens dipole allowed transitions (H\,I, He\,I, Na\,I, Ca\,II), the intercombination line Mg\,I~4571~\AA{}, and more than three tens forbidden lines of different atoms (O, N) and ions (Fe$^+$, S$^+$, N$^+$, Ca$^+$, Ni$^+$, O$^+$, O$^{+2}$).

\begin{figure}
\includegraphics[width=\columnwidth]{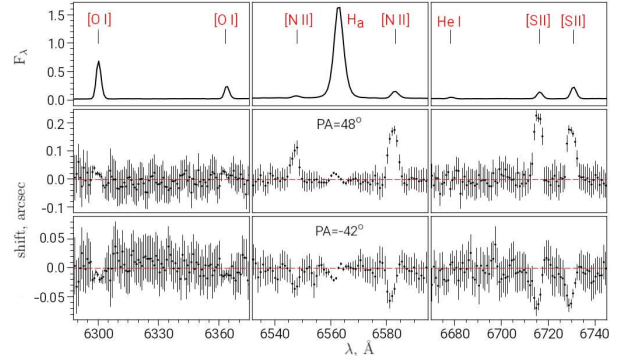}
  \caption{Spectro-astrometry for ZZ~Tau~IRS. Upper panel -- some sections of the spectrum with strong emission lines obtained in the {\lq}bright{\rq} state. The flux  $F_\lambda$ is in arbitrary units. The shifts of photocenter of the lines shown above relative to nearby continuum are shown for slit orientations PA$=48\degr$ (middle panel) and PA$=-42\degr$ (bottom panel).
  A positive shift along the slit corresponds to the north-east direction from the star at PA$=48\degr$ and to the north-west direction at PA$=-42\degr.$ It can be seen that the displacement of the photocenter of the [S\,II] and [N\,II] lines is significantly larger than that of H$\alpha$ and [O\,I] lines.}
 \label{fig:spectro-astrom}
\end{figure}

We used the TDS spectra of the star observed on rJD$=59503.55$ (slit orientation PA$=48\degr)$ and 59504.54 (PA$=-42\degr)$ to measure photocentre positions of the strongest emission lines relative to the adjacent continuum \citep{Whelan-2008} in the red channel, where the S/N ratio is high enough. The results of spectro-astrometric measurements are presented in Fig.~\ref{fig:spectro-astrom}, where one can see that the photocentre of these lines is shifted relative to the nearby stellar continuum in both spectra, indicating that the H$\alpha$ and forbidden lines originate in a significantly extended region. It also can be seen that the shift of [S\,II] and [N\,II] ion lines is 3-4 times larger than that of H$\alpha$ and [O\,I] atomic lines. 

\begin{figure}
\includegraphics[width=\columnwidth]{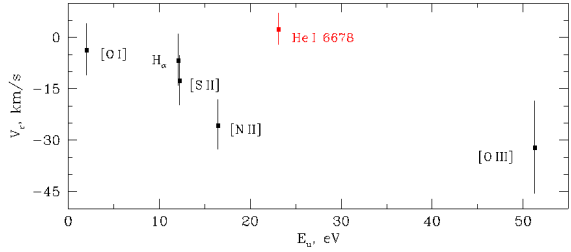}
\caption{The relation between radial velocity (averaged over all observed spectra) and the energy of upper level of respective transition measured from the ground state of the corresponding {\it atom} for the [O\,I]~$6300+6363$, H$\alpha$, [S\,II]~$6716+6731$, [N\,II]~$6548+6583$, [O\,III]~5007~\AA{} (black squares) and He\,I~6678~\AA{} (red square) lines. }
 \label{fig:Vr-TDS}
\end{figure}

Radial velocities $V_{\text r}$ for these two groups of lines are also different. This is clearly seen from Fig.~\ref{fig:Vr-TDS}, which shows the $V_{\text r}$ values (relative to the star), averaged for each line over all TDS spectra from Table~\ref{tab:spectra}, as a function of line upper level energy $E_{\text u}$ measured from the ground state of respective atom\footnote{
For example, in the case of the [O\,III]~5007~\AA{} line, $E_{\text u}$ is a sum of O and O$^+$ ionization potentials plus excitation energy of the ${}^1{\text D}_2$ level.}.
In the terms of the two-component model, the observed $V_{\text r}$ vs. $E_{\text u}$ dependence means that the higher $E_{\text u}$ of a line, the larger the contribution of HV component (jet) to its profile. Note in this regards that the [O\,III]~5007~\AA{} line, which has the largest blue shift, has been never observed in the disc wind of CTTSs, but it is present in the spectra of some Herbig-Haro objects, i.e. in jets \citep{Dopita-2017}. 

An important exception to the $V_{\text{r}} (E_{\text{u}})$ relation is the He\,I~6678~\AA{} line which has large enough $E_{\text u},$ but near-zero $V_{\text r}$. We believe this to be indicative that the line originates near the stellar surface in an accretion shock. Recall in this regard that He\,I~6678~\AA{}, unlike other emissions in the 2023 spectrum, has a symmetric one-component profile.

\section{Discussion}
\label{sect:discuss}

\subsection{Geometry of the outflow and outflow cavity}
 \label{sect:outflow -geometry}

\begin{figure}
\includegraphics[width=\columnwidth]{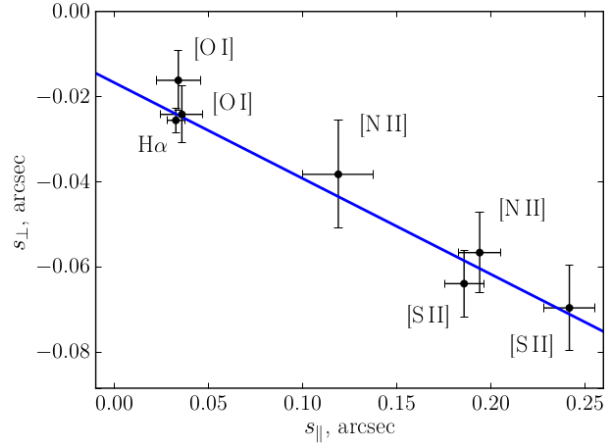}
\caption{The relation between spectro-astrometric shifts of the strongest lines photocentres. The shifts were derived from the red-channel TDS spectra obtained with the slit being aligned along PA$=-42\degr$ and PA$=48\degr$. }
 \label{fig:jet-orientation}
\end{figure}

As can be seen from Fig.~\ref{fig:jet-orientation}, the spectro-astrometric shifts of the line photocentres $s_\parallel$ (with the slit positioned at PA$=48$\degr) and $s_\perp$ (at PA$=-42$\degr) correlate. Using a generalized least-squares method \citep{astroMLText}, we derived the coefficient $k=-0.23 \pm 0.05$ for the linear function fitting the $s_\perp (s_\parallel)$ relation. In our opinion, the $s_\perp$ -- $s_\parallel$ correlation is due to the fact that there is a difference $\gamma$ between the direction of gas motion PA$_{\text j}$ and PA$=48${\degr}. In this case, $k=\tan \gamma$ and PA$_{\text j}=48\degr - \gamma = 61 \pm 3$\degr, if we take into account the sign of displacement from the star for different slit positions -- see the description for Fig.~\ref{fig:spectro-astrom}.       

%

\begin{figure}
\includegraphics[width=\columnwidth]{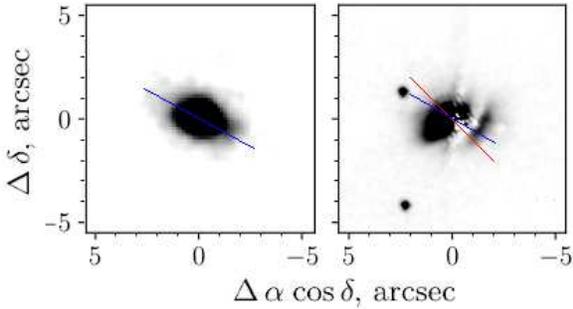}
\caption{The image of the vicinity of ZZ~Tau~IRS obtained by subtracting images in the [S\,II]~$6716+6731$ lines and nearby continuum (left panel) and the image of the same region obtained by the HST camera WFC3/IR in the F160W filter (right panel). In both panels, the blue line corresponds to the direction of the outflow axis (PA$=61$\degr) found from spectro-astrometry, and the red line in the right panel (PA$=45$\degr) shows the direction of the minor axis of the dust disc image according to \citet{Hashimoto-2021}. See text for details.}
 \label{fig:jet-disk}
\end{figure}

Subtracting the direct images obtained with the [S\,II]~$6716+6731$ and [S\,II]rc filters, \citet{Dodin-2023} found that ZZ~Tau~IRS has an elongated oval shape. As can be seen from the left panel of Fig.~\ref{fig:jet-disk}, the blue straight line with PA$=61$\degr runs approximately along the major axis of this oval, thereby confirming our conclusion that the gas responsible for line emission outflows along the direction PA$=61 \pm 3$\degr (outflow axis).

The HST image of ZZ~Tau~IRS in the NIR band (see Sect.\ref{sect:observation}) is shown in the right panel of Fig.~\ref{fig:jet-disk}. Judging by the size of two field stars in the left half of the frame, the image of ZZ~Tau~IRS has high spatial resolution, so one can identify an arc-like white gap between two dark {\lq}halves{\rq} of the image with the shadow from the disc -- compare, e.g., with Fig.~2 of \citet{Habel-2021}. The outflow axis (a blue straight line plotted over the HST image)
 seems to pass through about the middle of this arc, what means that the outflow axis is approximately perpendicular to the disc. 

The contribution of line emission in the 1.5--1.7~$\mu m$ range is small \citep{Kounkel-2019}, so the spatial extent of ZZ~Tau~IRS in the F160W filter results from continuum scattering by circumstellar dust. We conclude therefore, that the [S\,II] image traces line emitting region, i.e. the wind and the jet, whereas the F160W image contains information about the spatial distribution of the dust which scatters the light of the central star. As can be seen from Fig.~\ref{fig:jet-disk} these images have different shapes: unlike the [S\,II] image, the HST one is clearly asymmetric with respect to the outflow axis, it is more extended in the south-east direction along PA$\approx 150\degr$. 

The non-axisymmetric distribution of scattering matter in the shell is a plausible explanation of the abnormal  polarization angle -- $I$ brightness relation observed for ZZ~Tau~IRS\footnote{The contribution of emission lines is small in the $I$ band as follows from the Keck spectrum of the star.}.
We mean the following. The orientation of polarization vector $\theta_{\text{I}}$ in the {\lq}bright state{\rq}, indicates that scattering occurs on circumstellar dusty shell elongated along the rotation axis of the disc rather than on disc itself (Sect.~\ref{sect:photopolarimetry}). Then one can expect \citep{Shulman-2022} that, as the star fades, the polarization angle approaches the direction, perpendicular to the outflow axis, i.e. to PA$\approx 151\degr$ in our case. The $\theta_{\text{I}}(I)$ relation was observed earlier, e.g. in RW~Aur~A \citep{Dodin-19}, but in the case of ZZ~Tau~IRS the change of $\theta_{\text{I}}$ with brightness occurs in another way (see the right panel of Fig.~\ref{fig:polariz-vs-mag}). It may result from non-axisymmetric distribution of scattering dust \citep{Grinin-1988, Grinin-1994b, Shulman-2022}. However, to substantiate this hypothesis quantitatively, more polarimetric data at different wavelengths are needed.

As follows from Fig.~\ref{fig:jet-disk}, the extention of the scattering region is larger than $1.5\arcsec,$ i.e. $\gtrsim 200$~au, if we adopt a distance of 130~pc \citep{Akeson-2019}. It is reasonable to assume that scattering takes place on the inner surface of a cavity created by disc wind and jet in the protostellar cloud from which ZZ~Tau~IRS was formed -- a so-called outflow cavity. As regard the asymmetry of this cavity, we should note that \citet{Hashimoto-2021} found a ring-like dusty disc around ZZ~Tau~IRS with a bright crescent-like structure at PA$=135$\degr. It is quite natural to expect a relation between azimuthal asymmetries of the disc and cavity. 

PA of the dust disc major axis is $\approx 135\degr$ \citep{Hashimoto-2021}. So, if the matter outflows along the rotation axis of the disc, then the outflow axis should be oriented at PA$_{\text{a}}=135-90=45\degr,$ which differs noticeably from the value we found, PA$_{\text{j}}=61\pm 3 \degr$, as clearly demonstrated by the blue and red lines in the right panel of Fig.~\ref{fig:jet-disk}. At the same time the Herbig-Haro object HH\,393 is located in the direction of PA$_{\text{HH}} \approx 52 \degr$ from ZZ~Tau~IRS (see, e.g. Fig.~1 in \citealt{Dodin-2023}). We have no enough data to unambiguously interpret these differences in positional angles PA$_{\text{a}},$ PA$_{\text{j}}$ and PA$_{\text{HH}},$ and therefore limit ourselves to the following qualitative remarks. 

To explain the asymmetry of the dust disc of ZZ~Tau~IRS \citet{Hashimoto-2021} assumed that the inner disc ($r\lesssim 0.4\arcsec$) of ZZ~Tau~IRS is misaligned relative to the outer one, presumably due to the presence of a giant planet on an inclined orbit. If so, one can expect the axis of the jet and/or disk wind to precess leading to the asymmetry of the outflow cavity. To test this hypothesis, additional {ALMA} observations with better flux sensitivity and spatial resolution are needed. Note in this regard that \citet{Hashimoto-2021} estimated the ZZ~Tau~IRS mass from the {ALMA} $^{12}$CO~J$=3 \rightarrow 2$ line observation assuming the radial velocity of the star is $V_{\text{r}}=+6.5$~km/s that is $\approx 12$~km/s less than the value we used -- see Sect.~\ref{sect:star-spectra}. 
  
\subsection{Physical parameters of the jet and accretion rate}
 \label{sect:jet-accretion-params}

As we note in Sect.~\ref{sect:star-spectra}, the spectro-astrometric shifts and radial velocities of ion ([S\,II], [N\,II]) and atomic ([O\,I]) lines are significantly different. The two-component outflow model by \citet{Kwan-Tademaru-1988} naturally explains this feature by noting that the ion lines are formed in the jet, and the [O\,I] lines are formed in the disc wind of ZZ~Tau~IRS.

As a simple model of the jet we consider a thin spherical segment (shell), each point of which moves radially outwards with the same velocity $V_0.$ The shell is confined inside the cone with an angle of $2\alpha$ at its vertex, i.e. at the star, and the cone axis is inclined at angle $i$ to the line of sight, so that $i>\alpha.$ In this case the observed profile of an optically thin line with local $\delta$-function profile, is

\begin{equation}
F(v) = \frac{1}{\alpha}
\arccos \left[ 
\frac{\cos \alpha}{{\left( 1-v^2 \right)}^{1/2} \sin i  - v \cos i } 
\right],
 \label{eq:jet-prof}
\end{equation}
where $V_{\text r}$ -- is radial velocity, $v=V_{\text r}/V_0$ and $-\cos \left( i-\alpha \right) \leqslant v \leqslant -\cos \left( i+\alpha \right)$ \citep{Kwan-Tademaru-1995}. Note that the profile described by Eq.~\ref{eq:jet-prof} does not depend on distance of the shell from the star and on its thickness, so this model can include several separate shells, which move radially outwards with the same velocity $V_0.$

This profile with $V_0=84$~km/s, $\alpha=14\degr,$ and $i=i_j=74\degr$ appears to fit quite well the HV components of forbidden lines, as can be seen in the upper right panel of Fig.~\ref{fig:keck-profs} with the [S\,II]~6731~\AA{} line. As the jet axis should be perpendicular to the innermost accretion disc, the derived value of $i_{\text j}$ agrees with the notion that the protoplanetary disc of ZZ~Tau~IRS is seen close to edge-on. 

Remember that \citet{Hashimoto-2021} derived the angle between the {\it dust} disc axis and the line of sight $i_{\text d}=60\degr,$ that is significantly smaller than $i_{\text{j}}.$ But the model we used to estimate $i_{\text{j}}$ is too simplified to state that the inner disc is inclined to the outer one. An argument in favour of this interpretation can be derived from the following considerations. The presence of the [O\,III]~5006.8~\AA{} line in the ZZ~Tau~IRS spectrum (see Fig.~\ref{fig:tds-average-spectrum}) means that the jet bow shock moves with velocity $V_{\text{sh}} > 75$~km/s with respect to the matter in front of it \citep{Dopita-2017}. On the other hand, the radial velocity of this line in the star rest frame is $V_{\text{r}}=32.1 \pm 11.6$~km/s (see Fig.~\ref{fig:Vr-TDS}), therefore the jet is inclined to the line of sight at $i_{\text{j}}=\arccos\left( V_{\text{r}}/V_{\text{sh}} \right)\approx 64 \pm 10\degr.$ But the Herbig-Haro objects are believed to be shocks in jets produced by interaction within the jet of fast moving gas with slower matter ejected previously \citep{Raga-1990}. In other words, we talk about a shock that propagates through the moving jet matter and therefore the shock velocity relative to the star can be much greater than $V_{\text{sh}}$, and then $i_{\text{j}}$ should be significantly larger than the value we found.      

If the shock front velocity $V_{\text sh}$ is $>75$~km/s, then the post-shock temperature is $\gtrsim 10^5$~K \citep{Dopita-2017}. Thus, the UV radiation observed by {GALEX} \citep{GALLEX-2017} in the FUV $(\lambda_c \approx 0.15~\mu m)$ and NUV $(\lambda_c \approx 0.23~\mu m)$ bands can well be connected with jet emission in the C\,IV~1550 and Mg\,II~2800 doublet lines rather than emission from the stellar chromosphere and/or accretion shock.

\begin{figure}
\includegraphics[width=\columnwidth]{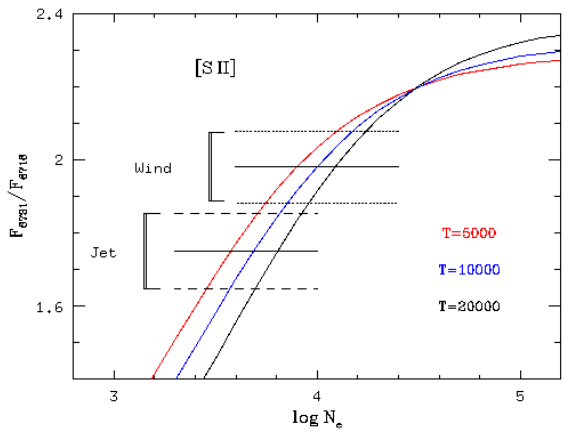}
\caption{Comparison of observed flux ratio of HV and LV components emission in the [S\,II]~6731 and 6716~\AA{} lines in ZZ~Tau~IRS spectrum with theoretical values from \citet{Nebulio-2015}.}
 \label{fig:S2-diagramm}
\end{figure}

We found that the flux ratios of HV and LV components emission in the [S\,II]~6731 and 6716~\AA{} lines are 1.98 and 1.75, respectively, with accuracy $\sim 5\,\%$. As follows from the diagnostic diagram based on data by \citet{Nebulio-2015} and presented in Fig.~\ref{fig:S2-diagramm}, it means that the electron number density in the regions where the [S\,II] lines are formed is $N_{\text e}^{\text{j}} \approx 5\times 10^3$~cm$^{-3}$ in the jet and approximately two times larger in the disc wind. 

Let us now consider the jet of ZZ~Tau~IRS. Judging by Fig.~\ref{fig:keck-profs} the gas radial velocity in the region emitting the [S\,II] lines exceeds 50~km~s$^{-1}$ in the star rest frame, so, the gas in this region moves outwards with the velocity $V_{\text{j}}>50$~km/s. A typical radius of a CTTS jet $R_{\text j}$ is $\gtrsim 10$~au \citep{Cabrit-2007}, and the hydrogen ionization degree in the [S\,II] line formation region behind the shock is usually $\sim 0.1-0.2$ \citep{Cabrit-2007, Melnikov-09}. Then we can estimate the jet mass loss rate for ZZ~Tau~IRS as follows:   

\begin{equation}
\dot M_{\text j} \approx \pi R_{\text{j}}^2 V_{\text{j}} m_{\text{H}} N_{\text{e}}^{\text{j}}   
{\left( \frac{N_{\text{H}}}{N_{\text{e}}} \right)}_{\text{j}} \ga 5\times 10^{-10}~M_\odot/{\mbox{yr}},
 \label{eq:Mdot-jet}
\end{equation}
where $m_{\text H}$ is the mass of hydrogen atom.

The accretion rate for CTTSs $\dot M_{\text acc}$ is typically an order of magnitude larger than the mass loss rate \citep{Cabrit-2007}. If it is true for ZZ~Tau~IRS, then its accretion rate $\dot M_{\text acc}$ is $>5\times 10^{-9}$~M$_\odot$~yr$^{-1},$ which is too large for a star with mass M$_* < 0.3$~M$_\odot$ (see Introduction): the statistical relation $\log \dot M_{\text acc}$ vs. $\log M_*$ with a standard deviation of 0.4 \citep{Alcala-2014}\footnote{See also \citet{Somigliana-2022} and references therein.} 
predicts an accretion rate almost an order of magnitude smaller than we found for ZZ~Tau~IRS. Thus, with all the uncertainty discussed above, we conclude that ZZ~Tau~IRS is an unusually active accretor. 

\subsection{The nature of ZZ~Tau~IRS variability}
 \label{sect:activity}

We believe that the strong photometric and spectral variability of ZZ~Tau~IRS is due to non-stationary disk accretion. The following arguments can be given in favor of this statement.

The colour-magnitude dependence observed for UXORs in visible and NIR bands is induced by the decreasing of scattering and absorption efficiency of dust grains with increasing wavelength. In the case of ZZ~Tau~IRS, this can explain why the star is {\lq}bluer{\rq} in the visible bands and {\lq}redder{\rq} in the NIR bands when fainter in the $I$ band (Fig.\ref{fig:colors}). However, this circumstance alone does not explain why the NIR brightness of ZZ~Tau~IRS increased when its brightness in visible band decreased. This behaviour is not typical for UXORs, though \citet{Shenavrin-2016} observed an increase of NIR brightness together with $V$-band fading for RR~Tau and CV~Cep. Besides, during a very deep $(\Delta V >10^m)$ eclipse of RW~Aur~A, a clear correlation between $V$ and $JHK$ magnitudes was observed, but it broke down for the $V$ and $K$ light at $\Delta V > 4$ -- see Fig.~5 in \citet{Dodin-19}. \citet{Shenavrin-2019} think that in these cases there is an increase of the emission from gas and dust at $\lambda > 1~\mu m$ coming from the inner parts of the accretion disc and/or dusty disc wind. One can assume that it is caused by the growing of the accretion rate resulting in the heating of the disc regions nearest to the star and in increase in disc wind.   

Let us now consider the cause of spectral variability. The equivalents widths of emission lines grew as the visible brightness of ZZ~Tau~IRS declined. For instance, the EWs of H$\alpha$ and [S\,II]~6731~\AA{} were 160 и 22~\AA{}, respectively, in the {\lq}bright{\lq} state (rJD$=59504.54,$ $I=15\fm39)$, and increased up to 376 and 79~\AA{}, respectively, when the star was near its minimum light (rJD$=59955.27,$ $I=16\fm52)$. At the same time, the monochromatic continuum fluxes $F_\lambda$ measured all over the TDS spectrum varied in the same way as the $V$ and $R$ light (Fig.~\ref{fig:colors}), i.e. correlated with the $I$ brightness.  

\begin{figure}
\includegraphics[width=\columnwidth]{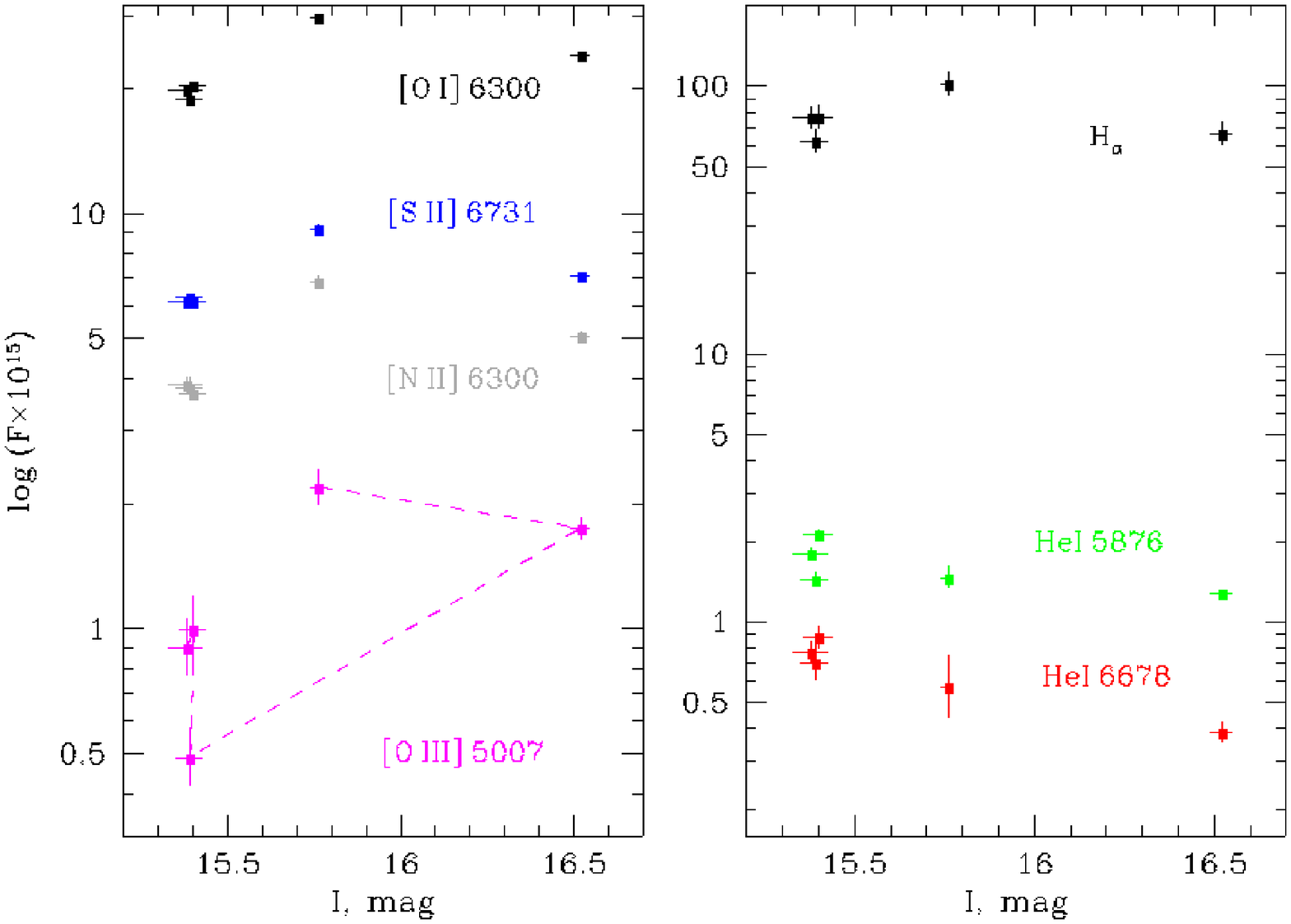}
\caption{The dependence of emission line fluxes $F$ (erg/s/cm$^2$) on the $I$ brightness for ZZ~Tau~IRS. The dashed line connects measurements that follow each other in time.}
 \label{fig:FvsI}
\end{figure}

The evolution of line {\it fluxes} is more complicated. As seen in Fig.~\ref{fig:FvsI}, when the star transitioned from the {\lq}bright{\lq} state to {\lq}faint{\lq} state, the forbidden line fluxes $F$ increased, the He\,I flux decreased, and the H$\alpha$ flux barely changed (right panel). To provide a quantitative description of spectral variability, we need to know how the profiles and fluxes of HV and LV components vary individually that is impossible because of low resolution of our TDS spectra. Instead we propose the following qualitative explanation for the observed relation $F(I)$ considering that each of the He\,I~6678~\AA{} and [O\,III]~5007~\AA{} lines has only one component -- LV and HV, respectively (see Section~\ref{sect:star-spectra}).     

If the He\,I~6678~\AA{} line is emitted by the accretion shock near the surface of ZZ~Tau~IRS, then the decrease of its flux with the star fading relates to an increase of optical depth in the dusty disc wind. This explanation is probably suitable for the relation $F(I)$ in the case of He\,I~5876~\AA{}, too. 

On the other hand, if the [O\,III]~5007~\AA{} line originates only in the jet, then the increase in its flux implies that the transition of ZZ~Tau~IRS from the {\lq}bright{\lq} state to the {\lq}faint{\lq} state was accompanied by the emergence of a new Herbig-Haro object in addition to the one already existing. We can present two arguments in favour of this interpretation. First, when the $I$ brightness of the star started to increase in the fall of 2023, the [O\,III]~5007~\AA{} line flux appeared even larger then in the {\lq}faint{\lq} state. Second, the deep eclipse of ZZ~Tau~IRS by dusty wind \citep{Dodin-19} was also accompanied by the emergence of a new Herbig-Haro object \citep{Takami-2023}.

Other forbidden lines demonstrate nearly the same relation $F(I)$, which must have the similar explanation. In our opinion, a smaller amplitude of flux variability of these lines is due to the contribution of the LV component that forms in the disc wind. As regard H$\alpha$, we should take into account that it forms not only in the wind and/or jet, but also in the accretion shock.  

So, the characteristics of line flux variability indicate significant changes in intensity of not only disc wind but of collimated outflow, too, that in the case of CTTSs is unambiguously related to variations in accretion rate \citep{Hartmann-2016}.


\section{Conclusion}
\label{sect:concludion}

It follows from our observations that photometric and polarimetric variability of ZZ~Tau~IRS results from eclipses of the star by circumstellar dust clouds. In other words, the star belongs to UX~Ori type stars, and as far as we know, it has the latest spectral class among them. The nature of eclipsing dust clouds may be different \citep{Ansdell-2020}, but our spectroscopic analysis shows that in the case of ZZ~Tau~IRS we deal with inhomogeneities in the dusty disc wind.

The presence of disc wind allows us to make the following conclusions about ZZ~Tau~IRS.
\begin{enumerate}
\item Abnormally large EWs of emission lines in the ZZ~Tau~IRS spectrum are usually explained by viewing the star through dusty upper layers of a flared protoplanetary disc, whereas the emission lines originate in the jet above the disc \citep{White-Hillenbrand-2004}. In other words, it is supposed that the disc axis is nearly perpendicular to the line of sight, and therefore, \citet{Hashimoto-2021} assumed that the outer disc is inclined to the inner one at $\Delta i \approx 30\degr.$ But if the star is  eclipsed not by a hydrostatic equilibrium disc atmosphere but by dusty disc wind, then $\Delta i$ may be much less.   

\item \citet{Hashimoto-2021, Hashimoto-2022} found out that the dust grains in the outer disc of ZZ~Tau~IRS are much larger than in the interstellar medium, and in the azimuthal crescent-like inhomogeneity the grain size exceeds 1~mm. But the magnitude-colour relation for the visible and NIR regions indicates the domination of grains with size $\lesssim 1~\mu m$ in the dusty wind. This means that either the grains grow faster in the outer disc than in the inner disc or the wind blows away predominantly small dust grains.   
 
\item The azimuthal inhomogeneity of the disc wind is responsible for the emergence of an asymmetric cavity inside the parental protostellar cloud, and also for the irregularity in variation of polarization angle and degree in the $I$ band. It seems reasonable to assume that the wind asymmetry is somehow related to the presence of the azimuthal inhomogeneity in the protoplanetary stellar disc.

\end{enumerate}

Our data for ZZ~Tau~IRS suggest the presence of not only a dusty disc wind but a jet, too. Judging by spectra obtained in February 2003, the bow jet shock moved with a velocity $V_{\text{j}}>75$~km/s, and the jet mass-loss rate $\dot M_{\text{j}}$ exceeded $5\times10^{-10}$~M$_\odot$/yr. In 2022--2023 the star transition from the {\lq}bright{\lq} to the {\lq}faint{\lq} state caused by an increase in intensity of dusty disc wind was accompanied by the emergence of a new collimated outflow.

The variation in the intensity of the disc wind and jet observed in 2021--2023 is likely due to non-stationary accretion of the protoplanetary disc matter. Episodes of increasing accretion rate must have occurred previously: in the end of 80s -- beginning of 90s of the last century the star was 2--3 times brighter in the visible and NIR regions than in 2021--2023. Besides, the star brightness at 3--5~$\mu m$ was higher in March 2023 ({\lq}faint{\lq} state) than in 2010. If we take into account the accretion rate estimate derived from the 2003 spectrum ($\dot M_{\text{acc}} > 5 \times 10^{-9}$~M$_\odot$/yr), we can state that ZZ~Tau~IRS is a very actively accreting young star with powerful outflow of matter.  
 
We should note in this context that \citet{Dodin-2023} detected motion of circumstellar matter with velocities of $\sim 50$~km/s on the scale of several thousand au in the neighbourhood of ZZ~Tau and ZZ~Tau~IRS. As the accretion and outflow rates are an order of magnitude greater in the case of ZZ~Tau~IRS than the similar quantities for the components of the ZZ~Tau\,AB binary system \citep{Belinski-2022}, it is reasonable to relate the detected motion of circumstellar matter with the ZZ~Tau~IRS activity. 
 


\section*{Add made at proof}
\label{sect:add}
 
\citet{Hashimoto-2024} performed a more thorough analysis of the HST
$(\lambda=1.6$~$\mu$m) image of ZZ~Tau~IRS disk used by us, and also found
that the inner regions of the disk are inclined relative to the outer ones by $15\degr.$
     
\begin{acknowledgments}
We are grateful to the staff members at the CMO SAI MSU for their cheerful assistance with the observations. We acknowledge with gratitude that we used the data from the SIMBAD database (CDS, Strasbourg, France), Astrophysics Data System (NASA, USA), NIST Atomic Spectra Database (https://www.nist.gov/pml/atomic-spectra-database) and The atomic line list v2.04 (https://linelist.pa.uky.edu/atomic/).
\end{acknowledgments}

\section*{Financial support}

A.\,V.\,Dodin (observations, reduction, interpretation) and B.\,S.\,Safonov (polarimetric observations and reduction) acknowledge support from RScF grant 23-12-00092. The results were acquired on equipment purchased through the M.~V.~Lomonosov Moscow State University Program of Development.  

\section*{Conflict of interest}
The authors declare that there is no conflict of interest.

\bibliographystyle{aspb1}
\bibliography{sample}

\end{document}